\documentclass[onecolumn]{revtex4}
\usepackage[usenames]{color}
\usepackage{graphicx}
\usepackage{bm}
\usepackage{ulem}

\begin{document}

\title{Detachment of semiflexible polymer chains from a substrate\\
- a Molecular Dynamics investigation}

\author{J. Paturej$^{1,2}$, A. Erbas$^{3}$, A.
Milchev$^{4}$, and V.G. Rostiashvili$^5$}
\affiliation{
$^1$ Leibniz-Institut of Polymer Research Dresden, 01069 Dresden, Germany\\
$^2$ Institute of Physics, University of Szczecin, Wielkopolska 15,
70451 Szczecin, Poland\\
$^3$  Department of Materials Science and Engineering, Northwestern University,
Evanston, IL 60208, USA \\
$^4$ Institute for Physical Chemistry Bulgarian Academy of Sciences, 1113 Sofia,
Bulgaria\\
$^5$ Max-Planck-Institute for Polymer Research, Ackermannweg 10, 55128 Mainz,
Germany}

\begin{abstract}
Using Molecular Dynamics simulations, we study the force-induced detachment of a
coarse-grained model polymer chain from an adhesive substrate. One of the chain
ends is thereby pulled at constant speed off the attractive substrate and the
resulting saw-tooth profile of the measured mean force $\langle f \rangle$ vs
height $D$ of the end-segment over the plane is analyzed for a broad variety of
parameters. It is shown that the observed characteristic oscillations in the
$\langle f \rangle$-$D$ profile depend on the bending and not on the torsional
stiffness of the detached chains. Allowing for the presence of hydrodynamic
interactions (HI) in a setup with explicit solvent and DPD-thermostat, rather
than the case of Langevin thermostat, one finds that HI have little effect
on
the  $\langle f \rangle$-$D$ profile. Also the change of substrate affinity with
respect to the solvent from solvophilic to solvophobic is found to play
negligible role in the desorption process. In contrast, a changing ratio
$\epsilon_s^B / \epsilon_s^A$ of the binding energies of $A$- and $B$-segments
in the detachment of an $AB$-copolymer from adhesive surface strongly changes
the $\langle f \rangle$-$D$ profile whereby the $B$-spikes vanish when
$\epsilon_s^B / \epsilon_s^A < 0.15$.
Eventually, performing an atomistic simulation of(bio)-polymers,
we demonstrate that the simulation results, derived from our
coarse-grained model, comply favorably with those from the all-atom simulation.
\end{abstract}

\maketitle	

\section{Introduction} \label{sec_Intro}

During the last decade a rapid development of the so-called single molecule
dynamic force spectroscopy (SMDFS) has enabled the direct observation of the
chemical dissociation (e.g., base-pare binding in DNA, or ligand-receptor
interaction in proteins) initiated by an external time-dependent force in the
pico-Newton range \cite{Merkel,Ritort,Franco}. Theoretical interpretation of
SMDFS for a single bond rupture has been suggested by Bell \cite{Bell},
and developed by Evans \cite{Evans_1,Evans_2}. The Bell-Evans (BE) approach is
built upon an Arrhenius relationship which describes the bond rupture rate
(``off''-rate) subject to a time-dependent force, $k_{\rm off} = \kappa_0 \exp (
x_{\beta} f/k_BT)$,  where $\kappa_0$ denotes the rupture rate in the absence of
applied force, $f$ is the applied force per bond, and $x_{\beta}$ is the
coordinate where the activation barrier is located. Here and in what follows,
$T$ denotes the temperature, and $k_B$ is the Boltzmann constant. In other
words, the effective activation energy is represented as a linear function of
the force, $E_{\rm b} (f) = E_{\rm b}^{(0)} - x_{\beta} f$. Under the condition
of fixed loading rate, $f = {\mathcal R} t$, it could be shown that $f =
(k_BT/x_{\beta}) \ln (\mathcal{R} x_{\beta}/\kappa_0 k_BT)$, i.e., the
detachment force grows linearly with the logarithm of loading rate
$\mathcal{R}$. This relationship is usually referred to as the Bell-Evans model
and employed for the measurement of dynamic strength of molecular bonds, cells'
adhesion, and protein unfolding by means of an atomic force microscopy (AFM)
\cite{Baro}. However, for multiply bonded attachments the $f$ vs $\ln{
\mathcal{R}}$ relationship shows a non-linear behavior which might be related to
a more complicated cascade of activation barriers \cite{Merkel_1}.

A single-stranded DNA (ssDNA), strongly adsorbed on a graphite substrate,
represents an example of a multiply-bonded bio-assembly. The desorption of such
tethered DNA molecule, induced by the applied force, has been first studied by
Jagota {\it et al.}~\cite{Jagota_1,Jagota_2,Jagota_3,Jagota_4}. At equilibrium,
the macromolecule can be desorbed either by using the displacement of the chain
end over the adsorbing surface (and measuring the fluctuating force), or by
fixing the pulling force, applied to the chain end, and monitoring the mean
displacement of the end segment over the plane.  It has been shown (analytically
and by means of Brownian simulation) that in the displacement control (DC)
desorption, the average force $\langle f \rangle$ - displacement $D$ profile
exhibits a characteristic set of saw-tooth (force-spikes) oscillations,
corresponding to the underlying base sequence of the ssDNA
~\cite{Jagota_1,Jagota_2,Jagota_3,Jagota_4}.

When the displacement profile reaches a steady state, i.e., the desorbed
monomers are far away from both ends of the chain, each maximum in the
saw-tooth oscillations corresponds to an energy barrier that has to be overcome
in order to complete the monomer desorption. In a real system, this energy
barrier, $G_b$, is quite complex and composed of various energetic and entropic
contributions. For instance, interaction energy between monomers and the
surface, $G_{\mbox{\tiny surf}}$, conformational entropic contributions and
enthalpic energies of polymer chains $G_{\mbox{\tiny conf}}$, contributions due
to direct additive interactions and entropic effects of water molecules near the surface, or with
the chain  $G_{\mbox{\tiny sol}}$, etc. One may assume that these
energetic components can be decoupled,
and the overall energy barrier can be expressed as $G_b \approx G_{\mbox{\tiny
surf}}+G_{\mbox{\tiny conf}}+G_{\mbox{\tiny sol}}+G_{\mbox{\tiny other}}$, where
$G_{\mbox{\tiny other}}$ represents all contributions that cannot be accessed
by a course-grained (CG) simulation model (such as hydrogen bonding of water
molecules near/around the monomers).
In that case, using Molecular Dynamic (MD) simulations, effects of various contributions on the polymer-surface interactions as well as non-equilibrium single molecule experiments can be tackled systematically and in detail.

Recently we have revisited the detachment theory of a strongly adsorbed
macromolecule by making use of a free-energy-based stochastic equation (the so
called Onsager equation) approach, and by performing extensive Molecular Dynamic
(MD) simulations \cite{Paturej}. This study has confirmed the force-spikes
response under DC and also demonstrated how the saw-tooth profile is smeared out
with growing detachment velocity $v_c$ and increasing mass of the
AFM-cantilever. Moreover, we have shown that the average detachment force versus
detachment velocity $v_c$ relationship exhibits a \textit{nonlinear} behavior
when plotted in semilogarithmic coordinates. The presence of fluctuations in our
model enables, among other things,  to calculate the probability distribution
function (PDF) of the fluctuating force at the cantilever, measured at the
moment of ultimate detachment, which is an experimental observable in laboratory
studies.

In the present paper we extend and generalize our previous MD
simulations~\cite{Paturej} so as to probe systematically the influence of
various energy contributions to the desorption energy barrier $G_b$, more
precisely, on the resulting force $\langle f \rangle$ - displacement $D$
profiles. Using a CG model, the pairwise interactions between monomers can be
tuned to understand their influence on displacement profiles. Similarly, by
introducing torsional and dihedral harmonic potentials in addition to the
classic bead-spring potentials (further details on the simulations scheme will
be given below), effects of  bending and/or torsional stiffness of the polymer
backbone on the detachment behavior are examined. At this point we should
also note that the energy components forming the overall energy barrier $G_b$
can be in phase with each other along the reaction coordinate (distance above
the substrate in this case). Hence, their addition can make an energy minimum
between two consecutive maxima more shallow or deeper as we will see later in
this paper.

The paper is organized as follows. First, in Section~\ref{sec_CG} we examine
the role of hydrodynamic interactions within the context of external
force-driven polymer desorption by comparing the effect of Langevin- and DPD
thermostats  within our coarse-grained (CG) model. We also check the role of
substrate wettability and its impact on the $\langle f \rangle$-$D$ profile.
Then, the desorption of an alternating $A-B$ copolymer with different binding
energies of the $A$- and $B$-monomers to the substrate is examined. In addition,
the effect of  bending and/or torsional stiffness of the polymer backbone on the
detachment behavior is studied. Eventually, in Section \ref{sec_AllAtom} we
report on atomistic MD simulation of polypeptide detachment, 
using poly-glycin and poly-phenylalanine, adsorbed on a crystalline carbon
substrate, and compare it to the generic behavior of our coarse-grained model.
Our report ends with a brief summary, presented in Section \ref{sec_Summary}.

\section{Coarse-grained simulations} \label{sec_CG}

\subsection{Model}

Similar to our previous study \cite{Paturej}, simulations of a coarse-grained
model were carried out based on a generic bead-spring model of a flexible
polymer chain \cite{kg}, composed of $N$ monomers, connected by nonlinear bonds
along the polymer backbone. The bonded (two-body) interactions in the chain is
described by the Kremer-Grest \cite{kg} potential, $V^{\mbox{\tiny
KG}}(r)=V^{\mbox{\tiny FENE}}(r) + V^{\mbox{\tiny LJ}}(r)$ with the so-called
``finitely extensible nonlinear elastic'' (FENE) potential given by
\begin{equation}
V^{\mbox{\tiny FENE}}= -\frac 12 k r_0^2 \ln{\left[ 1 - \left(\frac
r{r_0}\right)^2 \right]}
\label{fene}
\end{equation}
The non-bonded interactions between monomers were taken into account by means of
the Lennard-Jones (LJ) potential, given by:
\begin{equation}
 V^{\mbox{\tiny LJ}}(r) = 4\epsilon\left[
(\sigma/ r)^{12} - (\sigma /r)^6 + 1/4
\right]\theta(r_c-r).
\label{wca}
\end{equation}
In Eqs.~(\ref{fene}) and (\ref{wca}), $r=|\mathbf r_{ij}|$ denotes the distance
between the center of  monomer (bead) $i$ and $j$, $r_c$ is the cutoff distance,
while the energy scale $\epsilon$ and the length scale $\sigma$ are chosen as
the units of energy and length, respectively. Accordingly, the remaining
parameters are fixed at the values $k=30\,\epsilon/\sigma^2$ and $r_0=1.5\,
\sigma$ \cite{kg}. In Eq.~(\ref{wca}) we have introduced the Heaviside step
function $\theta(x)=0$ or $1$ for $x < 0$ or $x \geq 0$. We performed
simulations with  short- and long-range cutoff: $r_c=2^{1/6}\, \sigma$ (purely
repulsive interaction between monomers), and  $r_c=2.5\, \sigma$ (monomer
attractions allowed at larger distances). In the course of the study, chain
bending stiffness $\kappa$ and torsional stiffness $\kappa_t$ were varied by introducing a
three-body,
\begin{equation}
V^b(\theta_{ijk}) = \kappa (\cos{\theta_{ijk}} -1)^2,
\label{stiff}
\end{equation}
and four-body interactions
\begin{equation}
V^t(\phi_{ijkl}) = \kappa_t(1+\cos{\phi_{ijkl}}),
\label{torsion}
\end{equation}
where  $\theta_{ijk}$ and $\phi_{ijkl}$ denote bending and dihedral angle formed
respectively by two and three successive bond vectors. In the CG-simulations two
kinds of substrates were considered. We employed structureless adsorbing surface
(with no friction in the lateral plane), modeled simply by a Lennard-Jones
potential acting with strength $\epsilon_s$ in the perpendicular $z$-direction,
$V^{\mbox{\tiny sub}}(z)=4\epsilon_s[(\sigma/z)^{12} - (\sigma/z)^6]$. In a
separate set of simulations, we introduced a rough surface composed of beads
which form triangular lattice and interact with monomers via Eq.~(\ref{wca}) in
order to take into account friction between polymer and substrate.

In our simulations we consider, as a rule, the case of strong adsorption
$\epsilon_s/k_BT=5$ and $20$ for the structureless surface, and
$\epsilon_s/k_BT=5$ in case of atomistic surface, with $T$ being the temperature
of the thermal bath which is described briefly below.


Temperature in our simulations was controlled by two different methods: (i) a
Langevin thermostat \cite{schneider}, and (II) by Dissipative Particle Dynamics
(DPD) thermostat \cite{dpd}. In both methods the dynamics of the chain is
obtained by solving the following set of equations of motion for the position
$\mathbf r_n=[x_n,y_n,z_n]$ of each bead in the chain,
\begin{equation}
m\ddot{\mathbf r}_n = \mathbf F_n^{\mbox{\tiny cons}} +
\mathbf F_n^{\mbox{\tiny D}} + \mathbf R_n \qquad
(1,\ldots,N)
\label{langevin}
\end{equation}
with $\mathbf F_n^{\mbox{\tiny cons}}$ being the total conservative force acting
on each polymer bead with mass $m=1$.

The influence of the solvent is split into slowly evolving viscous force and
rapidly fluctuating stochastic force. Thus, in Eq.(\ref{langevin}), $\mathbf
F_n^{\mbox{\tiny D}}$ and $\mathbf R_n$ denote respectively the dissipative and
random forces which are responsible for keeping the system at constant
temperature. The difference between Langevin and DPD thermostats lies is the
choice of these two forces.

In the Langevin thermostat, the dissipative force (drag force) is proportional
to particle's velocity $\mathbf F_n^{\mbox{\tiny D}}=-\gamma_{\mbox{\tiny
L}}\dot{\mathbf r}_n$, where $\gamma_{\mbox{\tiny L}}=0.5\, m\tau^{-1}$ is the friction
coefficient, and the time unit is $\tau=\sqrt{m\sigma^2/\epsilon}$. In addition,
the random force has a zero mean value and satisfies the fluctuation-dissipation
theorem, $\langle \mathbf R^{\alpha}_n(t) \mathbf R^{\beta}_m(t') \rangle
=2\gamma_{\mbox{\tiny L}} k_BT\delta_{\alpha\beta}\delta_{ij}\delta(t-t')$.
These two forces, the random and the frictional one, are balanced in order to
maintain the system temperature at the set value.

In contrast to the Langevin thermostat, in the DPD thermostat both dissipative
and random forces are applied as pairwise interactions, such that the sum of
these two forces acting on a given pair of particle in the system is zero. Thus,
in the DPD case the particle momentum is conserved, leading to correct
description of hydrodynamic interactions \cite{frenkel,soddemann}. The form of
dissipative and random forces in the case of DPD thermostat is the following:
$\mathbf F_n^{\mbox{\tiny D}} =\sum_{m(\neq n)} -\gamma_{\mbox{\tiny DPD}}
w^2(r_{nm}) (\mathbf r_{nm}/|\mathbf r_{nm}| \cdot \dot{\mathbf r}_{nm})\mathbf
r_{nm}/|\mathbf r_{nm}|$ and  $\mathbf R_n = \sum_{m(\neq n)}
\sqrt{2k_BT\gamma_{\mbox{\tiny DPD}}} w(r_{nm})\alpha_{nm} / \sqrt{dt}$, where
the weighting function $w$ is defined as $w(r) = 1- r/r_c$, $\gamma_{\mbox{\tiny
DPD}}=20\,m\tau^{-1}$ is the friction coefficient, and $\alpha$ is a Gaussian-distributed
random number with zero mean and variance equal to unity, whereas $dt$ stands
for the integration step.

In all CG-simulations the equations of motion were integrated using the MD
package LAMMPS \cite{lammps}. The solvent in our simulations was considered
either as being present implicitly via Langevin thermostat or
modeled explicitly by adding spherical particles with density
$\rho=0.86\,\sigma^{-3}$. In the latter case the difference between Langevin thermostat and
DPD thermostat was investigated.

The detachment of chains, composed of $N=20$, or $N=100$ monomers, was performed
as follows. In the initial state, the chains were completely adsorbed and
equilibrated on the surface. The macromolecule was then pulled perpendicular to
the adsorbing surface by a cantilever at constant velocity $\mathbf V
=[0,0,v_c]$. As a cantilever we used two beads connected by harmonic spring and
attached to one of the ends of the chain. The mass of the beads, forming the
cantilever, $m_c$, was set to $m_c=1$ whereas the equilibrium length of the
harmonic spring  was set to $0$ and the spring constant was chosen as
$k_c=50\epsilon/\sigma^2$. During the pulling simulations, the force $f(t)$ at
given height $D$ over the substrate was calculated from the instantaneous
harmonic linker extension $\Delta z_l(t)$, i.e., $f(t)=k_c\Delta z_l(t)$.

\subsection{Results} \label{sec_Results}

\subsubsection{Impact of chain properties on the $\langle f \rangle$-$D$
profile}

\begin{figure}[ht]
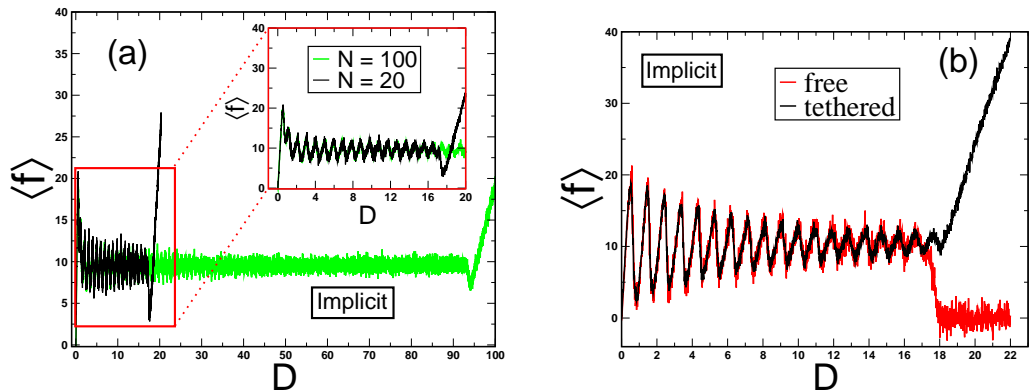

\begin{center}
\includegraphics[scale=0.25]{semiflexible_comparsion.eps}
\hspace{0.50cm}
\includegraphics[scale=0.25]{fvsD_nontet.eps}

\caption{Short vs long chain desorption in the case of implicit solvent. (a)
Mean detachment force $\langle f \rangle$ vs distance $D$ profile is shown for a
semi-flexible polymer chain (stiffness parameter $\kappa = 50$). Results pertain
to chains composed of respectively $N=20$, and $100$ monomers. For better
visibility, the onset of the desorption process is zoomed in the inset. (b)
Detachment of a chain whose end is {\it not} tethered to the substrate. In both
figures $\epsilon_s/k_BT=20$ and $v_c=10^{-3}\, \sigma/\tau$.}
\label{Finite_Size_Effect}
\end{center}
\end{figure}

The saw-tooth response of the pulling force $\langle f \rangle$, measured at any
fixed distance $D$ of the chain end above the adsorbing surface, is a
characteristic feature produced by the detachment of successive monomers along
the polymer backbone, cf. Figure \ref{Finite_Size_Effect}a. The observed steady
and steep increase of $\langle f \rangle$ after the characteristic last minimum,
which corresponds to detachment of the last monomer, is a hallmark of a chain,
tethered by one of its ends to the adsorbing substrate and pulled by the other
end monomer. It reflects the ultimate extension of the linker spring once all
beads, besides the tethered one, have lost contact with the adsorbing plane. In a real
 experiment, the chains which have to be detached from the adsorbing surface are
most probably not tethered, so the final $\langle f \rangle$-$D$ profile
exhibits instead a sharp drop of the force, which can be also found in our
simulations as shown in Figure~\ref{Finite_Size_Effect}b. This kind of behavior
has been observed before for fully flexible homopolymer chains by  Jagota {\it
et al.} \cite{Jagota_1,Jagota_2,Jagota_3,Jagota_4} and Paturej {\it et al.}
\cite{Paturej}.

Apparently, a semi-stiff tethered chain exhibits the same pattern, ({\it cf.}
Fig. \ref{Finite_Size_Effect}a), albeit the ultimate steep growth of the pulling
force $\langle f \rangle$ is preceded by a characteristic minimum extending over
the last few beads of the chain that precede the tethered bead. Obviously, the
last portion of the semi-rigid chain bends and {\it takes off as a whole},
whereby the length of this chain portion should depend on the chain stiffness.
No finite size effect can be detected on Fig. \ref{Finite_Size_Effect}a. Indeed,
it can be seen that, irrespective of the chain length (either $N=20$, or
$N=100$), the amplitude and position of the spikes remain insensitive to chain
length $N$. In addition, there is no significant difference between the
$\langle f \rangle$-$D$ profiles for polymer chains adsorbed on a smooth or
rough surface (not shown). The only difference is a slightly larger ($\approx
5\%$) amplitude of force $\langle f \rangle$ measured in the case of rough
surface.

\begin{figure}[ht]
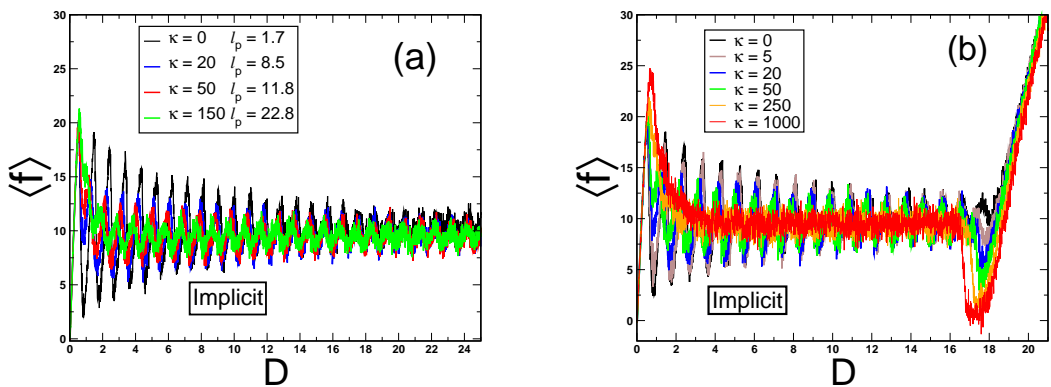

\begin{center}
\includegraphics[scale=0.25]{semiflexible.eps}
\hspace{1cm}
\includegraphics[scale=0.25]{semiflexible_N20.eps}
\caption{Force-distance diagram for semiflexible chains in implicit solvent.
Stiffness parameter $\kappa$ and persistence length $l_p$ are indicated  in the
legend: (a) Chain length $N=100$. With growing stiffness $\kappa$ the spikes
amplitude {\it decreases}. (b) Chain length $N=20$. The end of the detachment
process is marked by a force minimum which becomes more pronounced with growing
stiffness. Here $\epsilon_s/k_BT=20$ and $v_c=10^{-3}\, \sigma/\tau$.
}\label{Semiflexible_100_vs_20}
\end{center}
\end{figure}

Generally, one may speculate how the chain stiffness affects the $\langle f
\rangle$ vs $D$ diagram. The bending stiffness of the polymer backbone was
included in the simulation by allowing for the three-body bending potential,
$V_b(\theta)$, where $\theta$ is the angle between two consecutive bonds, cf.
Eq.~(\ref{stiff}). The stiffness parameter $\kappa$ determines the persistence
length of the chain, $l_p$, which is defined through the decay of bond angle
correlations \cite{Khokhlov}. Figure \ref{Semiflexible_100_vs_20} shows
simulation results for different stiffness parameter $\kappa$ and chain length
$N$.

\begin{figure}[ht]
\begin{center}
\includegraphics[scale=0.3]{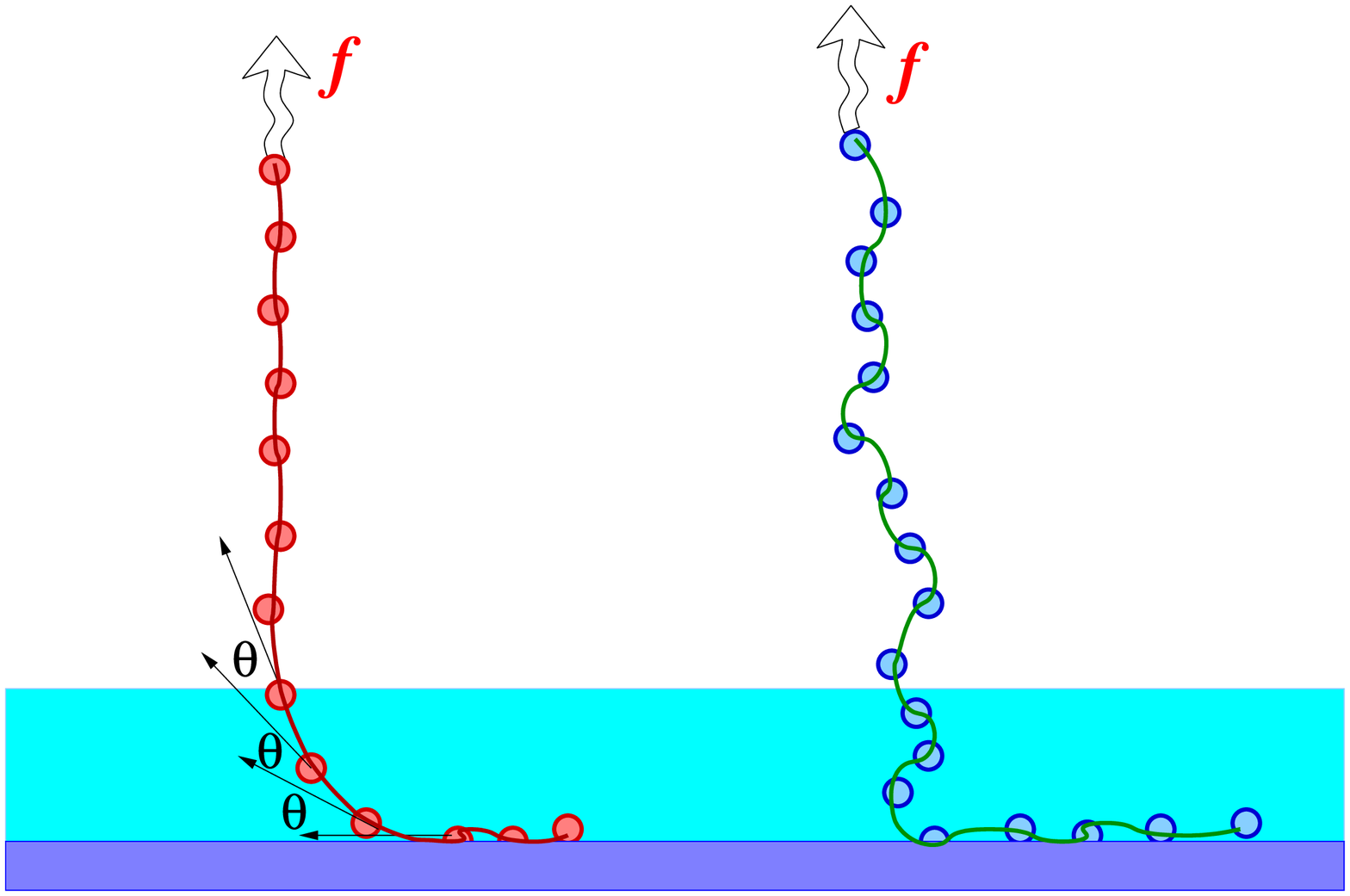}
\hspace{1.0cm}
\includegraphics[scale=0.15]{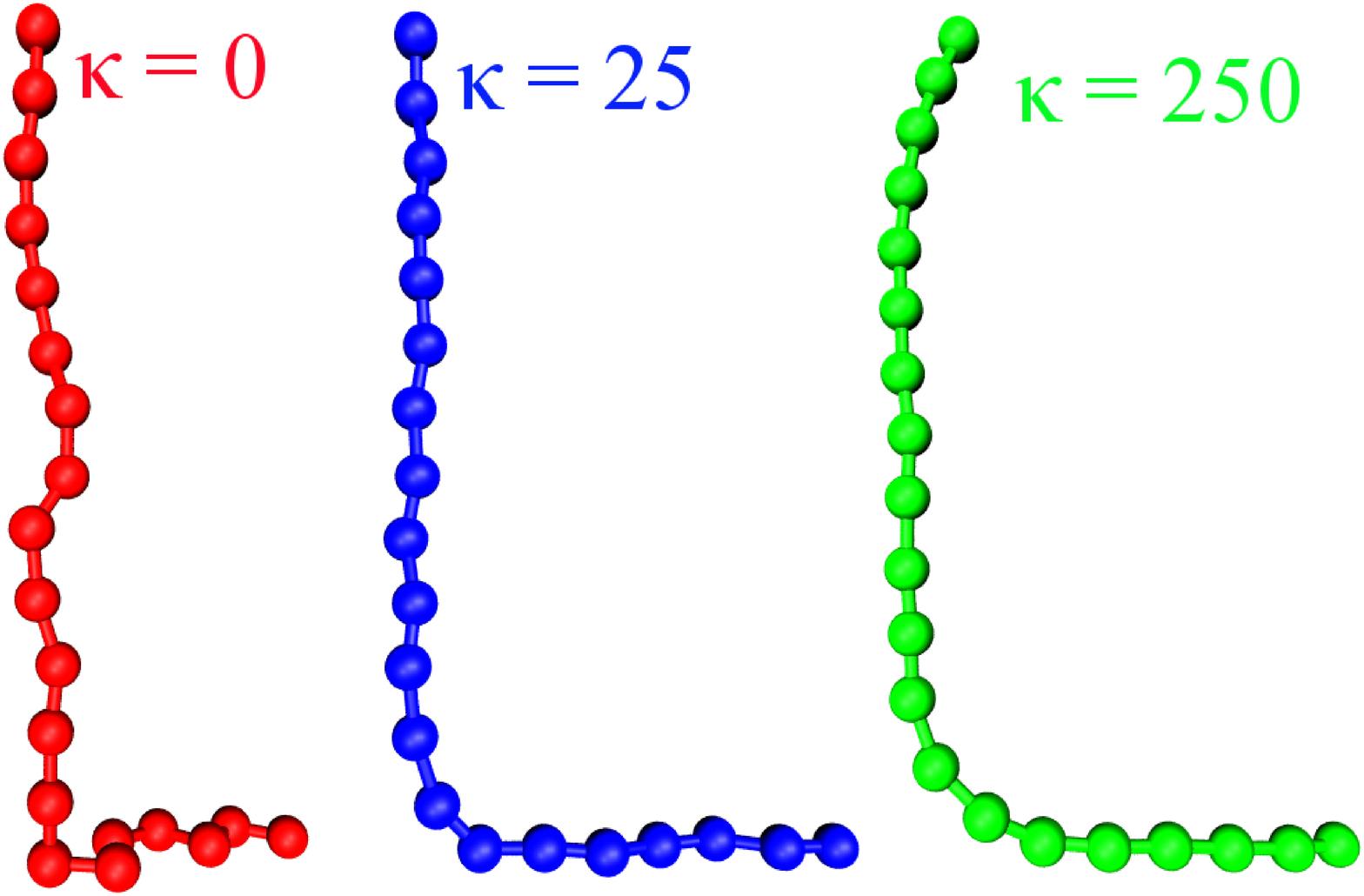}
\caption{(a) Schematic picture of a force-induced detachment of a semi-stiff
(left) and flexible (right) chain from a solid plane. The bending angle $\theta$
between successive bonds is indicated. Light-shaded area denotes the range of
the adsorption potential $V^{sub}(z)$. (b) Snapshots of partially detached
polymer chains  of different stiffness $\kappa$ as indicated. In each case 66\%
monomers were peeled off the substrate. }
\label{snapshots}
\end{center}
\begin{picture}(0,0)
 \put(-205,130){$\bf (a)$}
 \put(10,130){$\bf (b)$}
\end{picture}
\end{figure}

It can be seen that with growing stiffness, the amplitude of the spikes
decreases and also the spikes resolution deteriorates. Starting with
approximately $\kappa = 50$, the complete chain detachment is preceded by a
characteristic force minimum (in case of  tethered chains). Then, with growing
stiffness $\kappa$ this minimum, which marks the ultimate chain detachment,
occurs at smaller $D$, indicating that increasingly larger portions of the
polymer backbone are detached as a whole. The minimal detachment force thereby
also drops so that at $\kappa = 1000$ it approaches zero. However, for realistic
values of the stiffness $\kappa$, and not extremely short chains, $N \ge 10$,
the total energy, $E_{rod}$, needed for tearing off the polymer as a single
piece of rod from the adsorbing plane, $E_{rod} \propto N\epsilon_s$, would be
huge in comparison to the energy, $E_{arc}$, needed to tear off the same
semi-flexible chain bead by bead, $E_{arc} \approx n_m \epsilon_s +
(n_m-1)\kappa  \theta_b^4$, with  few beads $n_m$ that form an arc of the bended
portion of the chain backbone, Fig.~\ref{snapshots}. Once an arc is formed, no
further energy penalty will be needed to keep the chain bended during the rest
of the detachment process.

A rough estimate, using the potential $V^b(\theta)$, Eq.~(\ref{stiff}), with
$\theta = 10^o$ degrees  yields $\kappa \times 5.3 10^{-8}$ bending energy per
bond, or $E_{arc} \approx 10^{-5}\, k_BT$ for an arc encompassing $10$ bending
angles and $\kappa = 100\, k_BT$. In the same time such an arc can already reach
a height of $ 10\times \cos(10^o)\sigma \approx 1.74 \sigma$, where surface
adhesion is already dwindling. Therefore, the detachment of even rather stiff
chains instantaneously as a rod-like object should be ruled out and a $\langle f
\rangle$-$D$ profile of the type, shown in Figure~\ref{Semiflexible_100_vs_20},
is likely to be observed.

Owing to the creation of arc, more that one bead detach concertedly  and move
away from the adhesive surface beyond the range of adsorption. The neighboring
beads along the arc, $i$ and $j$, remain thereby at fixed mutual distance
$z_{ij} < \sigma$. The neighboring bonds slightly bend but do not stretch
significantly so that the length of the individual bond is close to the
unperturbed length yet much less than the maximal one, $r_0 = 1.5\sigma$, cf.
Eq.~(\ref{fene}). As a result, depending on chain stiffness and in contrast to
flexible chains, the monomers in a semi-stiff chain detach concertedly rather
than one by one which exerts a smearing effect on the saw-tooth diagram $\langle
f \rangle-D$, Fig.~\ref{Semiflexible_100_vs_20}a. Increasing bending stiffness
also decreases the magnitude of saw-tooth profile amplitude as the bonds
between neighboring monomers stretch less.

\begin{figure}[ht]
\begin{center}
\includegraphics[scale=0.3]{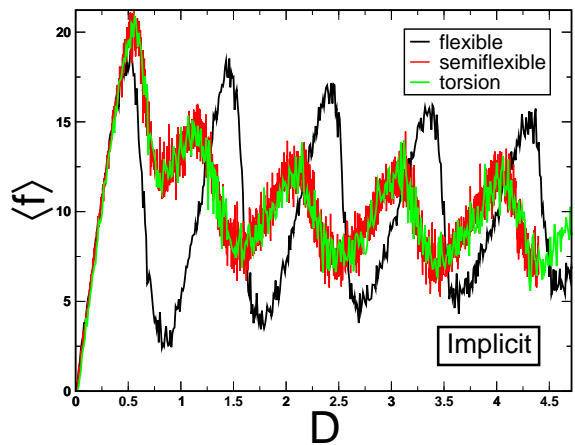}
\caption{Force $\langle f \rangle$ vs $D$ diagram for flexible, semiflexible and
torsional (angle and dihedral potential included) as indicated. Here $N=20$,
$\epsilon_s/k_BT=20$, $v_c=10^{-3}\, \sigma/\tau$, $\kappa=50\,k_BT$ and
$\kappa_t=\,k_BT$. } \label{Torsion}
\end{center}
\end{figure}
Realistic DNA-models usually include also a dihedral potential which is
responsible for the chain resistance to  torsion. We have used the dihedral
potential, $V_t(\phi)$, where $\phi$ is the dihedral angle and the torsion
constant is $\kappa_t = k_BT$, Eq.~(\ref{torsion}). The resulting $\langle f
\rangle$ vs. $D$ diagram for a chain with torsional and bending finite
stiffness, compared to the fully flexible, and semi-flexible ($\kappa = 50$)
chain models, is shown in Fig. \ref{Torsion}. It can be seen that while the
bending stiffness itself leads to a clear shift of the force oscillation
pattern, the resulting behavior practically does not change upon inclusion of a
dihedral potential.

\subsubsection{Effects of substrate adhesion on polymer detachment}

\begin{figure}[ht]
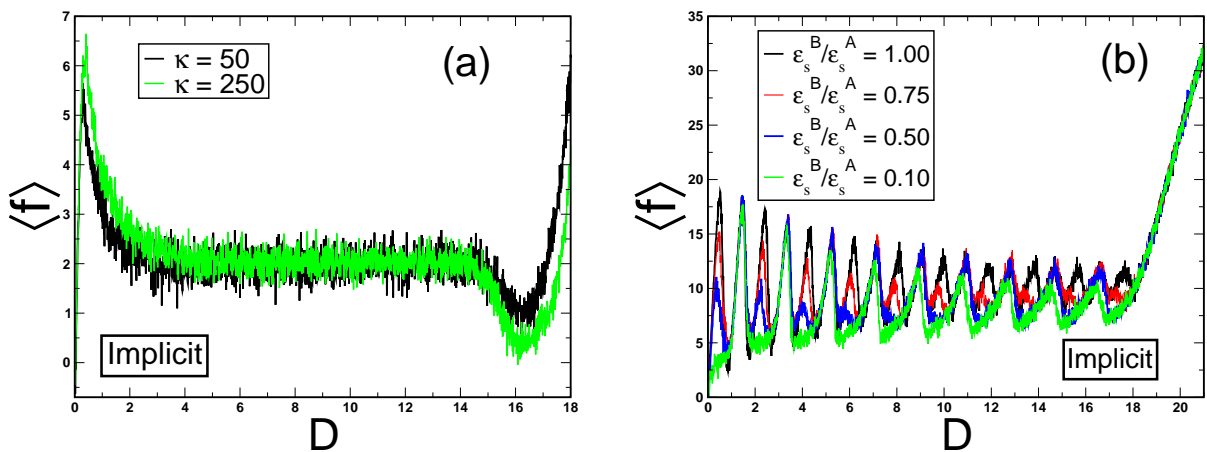

\begin{center}
\includegraphics[scale=0.3]{fvsD_semi_loweps.eps}
\hspace{0.5cm}
\includegraphics[scale=0.3]{copolymer.eps}
\caption{(a) Force $\langle f \rangle$ vs $D$ diagram of a homopolymer for
attraction strength of the surface $\epsilon_s = 5 k_BT$ and two different
values of the bending stiffness, $\kappa = 50,\; 250\,\,k_BT$. (b) The same for
alternating copolymers made of two types of monomers $A$ and $B$ which have
different binding energies to the substrate, $\epsilon_s^A$ and $\epsilon_s^B$,
respectively. Results are displayed for different ratios
$\epsilon_s^B/\epsilon_s^A$ as indicated in the legend. The case of
$\epsilon_s^B/\epsilon_s^A = 1$ corresponds to a homopolymer. The absolute value
of $\epsilon_s^A$ here is $20\,k_BT$. In both figures $N=20$  and $v_c=10^{-3}\,
\sigma/\tau$. } \label{Copolymer}
\end{center}
\end{figure}

It is to be expected that the strength of adhesion of the polymer chain to the
adsorbing surface will manifest itself in the recorded variation of desorption
force $\langle f \rangle$ with distance $D$. While in the previous graphs,
Figures~\ref{Finite_Size_Effect}-\ref{Torsion}, we focused on cases of strong
adsorption, $\epsilon_s = 20k_BT$, in Figure~\ref{Copolymer}a we present the
desorption profile for weak to moderate attraction of the chain by substrate
(in our model the threshold for adsorption $\epsilon_s^{crit} \approx 3 k_BT$).
Indeed, as indicated in Figure~\ref{Copolymer}a, at $\epsilon_s = 5 k_BT$, the
characteristic oscillations in the $\langle f \rangle$-$D$ profile virtually
vanish (apart from the statistical noise). Therefore, one may conclude
that the method of single chain detachment spectroscopy as a tool for
sequencing analysis could be used in cases of strong polymer - substrate
adhesion only.

For the objectives of sequencing, the legibility of the data, derived by this
method of force-induced detachment, must be examined for heterogeneous polymers
in particular. As an example, the result for an alternating ($A-B$)-copolymer
detachment is shown in Figure~\ref{Copolymer}b. Here monomers $A$ and $B$ have
different affinity to the  substrate (albeit the same mass $m_A = m_B$). The
detachment starts with a $B$-monomer (i.e., a monomer with a relatively smaller
affinity to the substrate). For a ratio of $\epsilon_s^B/\epsilon_s^A = 0.5$,
the alternating pattern of spikes can still be clearly seen. As $\epsilon_s^B$
gradually further declines, the set of force maxima, corresponding to the
desorption of $B$-monomers, decreases significantly in amplitude. Eventually,
for $\epsilon_s^B/\epsilon_s^A = 0.1$, the maxima corresponding to tearing-off
$B$-monomers turn into minima. Moreover, the latter effect is observed even at
higher values of the $\epsilon_s^B/\epsilon^A$-ratio once the first few
repeating units are been detached. So at larger height $D$ of the pulled chain
end (approximately starting from $D = 8 \sigma$), for the $B$-type monomer
desorption one observes local minima rather than peaks. Evidently, a correct
sequencing of heterogeneous macromolecules can be performed only in cases when
the affinity of the various building blocks is large in terms of  absolute values of binding energy $(>10\,k_BT)$
and the differences between values of binding energy should be significant.

\subsubsection{Implicit vs. explicit solvent}

So far we examined how the force-displacement profile of different polymer
chains reflects the properties of the chains and their interaction with the
adsorbing surface. It is of some interest to check whether the properties of the
surrounding medium, considered in the different simulation setups, might
influence the $\langle f \rangle$-$D$ profile too.

In many computational experiments, as e.g. in our previous publication
\cite{Paturej}, one takes the solvent only implicitly into account. The solvent
properties can be then varied to a limited amount only, for example, by changing
the friction coefficient $\gamma$ in Eqs.~(\ref{langevin}). In principle,
however, the presence of an explicit solvent might affect the course of
force-induced chain desorption due to hydrodynamic interactions (HI). To this
end we compare the $\langle f \rangle$ vs $D$ diagram, derived from simulations
of the same system when two different thermostats are used: (i) a Langevin
thermostat (i.e., with no HI), and (ii) a DPD thermostat. It is well known that
the latter allows for a correct hydrodynamic behavior \cite{frenkel,soddemann},
whereas the Langevin thermostat does not exhibit momentum conservation, and
therefore does not reproduce the proper hydrodynamic behavior. In the case of
explicit solvent, a chain is pulled in a Lennard-Jones liquid with liquid
monomer density $\rho = 0.86\sigma^{-3}$.
\begin{figure}[ht]
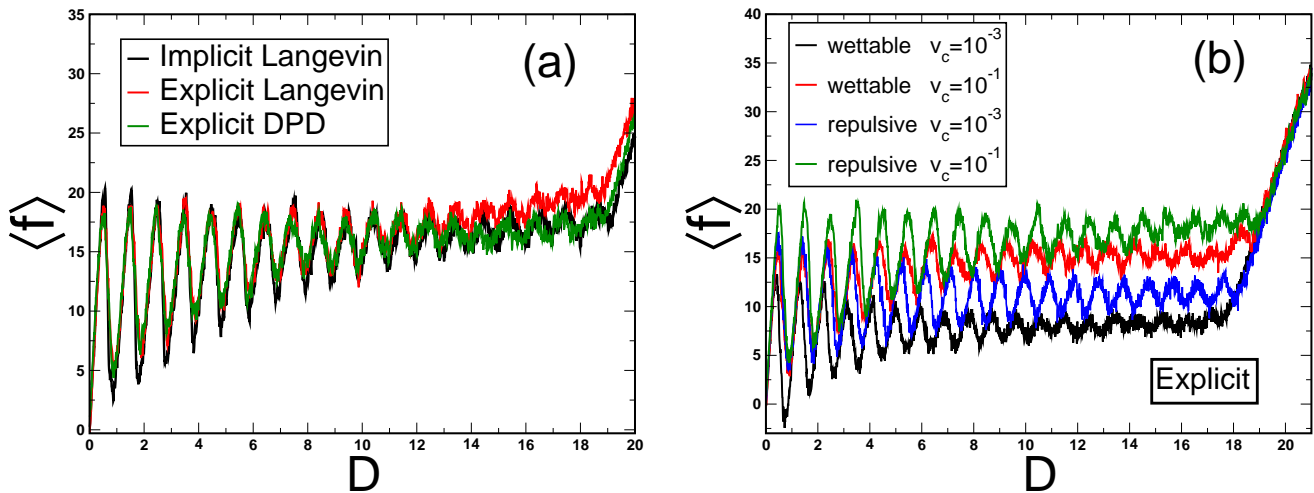

\begin{center}
\includegraphics[scale=0.33]{fvsD_fast_thermostats.eps}
\hspace{0.3cm}
\includegraphics[scale=0.33]{fvsD_selvsads.eps}
\caption{(a) Comparison of the force $\langle f \rangle$ vs $D$ diagram for
implicit and explicit solvents from simulations with Langevin, and DPD
thermostats, respectively. Results are presented for  fast pulling, $v_c =
10^{-1}\, \sigma/\tau$. (b) Impact of the substrate selectivity for the case of
explicit solvent and DPD thermostat. Force $\langle f \rangle$ vs. displacement
$D$ diagram for wettable and solvophobic substrates at two pulling velocities.
In both figures $N=20$ and $\epsilon_s/k_BT=20$.}
\label{Implicit_vs_explicit}
\end{center}
\end{figure}

It is evident from Figure~\ref{Implicit_vs_explicit}a, however, that there is no
tangible difference between these cases, which suggests that the hydrodynamic
interaction is largely irrelevant in detachment experiments. In fact, when
compared to the case of explicit solvent with no HI (Langevin thermostat), the
presence of HI (accounted for by DPD) leads to a slight decrease in the pulling
force, cf. Figure~\ref{Implicit_vs_explicit}a, at the end of the detachment
process. Evidently, this affects the detachment of the last beads only while the
main portion of the chain is sufficiently far away from the substrate. While
such an effect is completely missing for slow detachment with $v_c = 10^{-3}\,
\sigma/\tau$ (not shown here), for fast detachment, $v_c = 10^{-1}\,
\sigma/\tau$, it should actually be expected due to the solvent back-flow,
triggered by Stokes friction of the detached chain portion when the desorbed
chain eventually sets into motion. Moreover, due to confinement effects, HI
(being long-ranged) are screened \cite{Winkler} in the vicinity of the adhesive
wall, which explains why their presence is detectable only at sufficiently large
distance $D$ from the wall. Therefore, one can view the small decrease in
$\langle f \rangle$ in the case of explicit solvent as a typical manifestation
of the well-known difference between Rouse and Zimm dynamics of polymers.

\subsubsection{Substrate wettability}

We checked also to what extent the affinity of the adsorbing substrate with
respect to solvent plays a role. Basically, we distinguish between {\it
solvophobic} (repulsive) substrates, where the polymer chain is still attracted to the
surface whereas the solvent particles are repelled, and {\it wettable
substrates}, where both the polymer and the solvent are attracted to the
surface. Figure \ref{Implicit_vs_explicit}b indicates a systematic decrease in
the amplitude of the spikes for wettable substrates (as if the solvophobic
solvent effectively increases the chain adhesion to the surface), yet the
characteristic saw-tooth force vs distance profile remains qualitatively
unchanged. The spikes positions under conditions of good wetting are also
slightly shifted to lower values of $D$, i.e., the monomers detach more easily
(at somewhat lower height) as compared to the solvophobic case.
 This is  because as the relative attraction of solvent particles to the surface is increased, the solvent particles try to replace chain monomers and form a solvent layer on the surface, which in turn, facilitate the desorption of chain monomers.   In the context of protein-surface interactions, this effect is  referred to as \textit{Berg limit} and was also observed in the  simulations of biopolymers on various hydrophobic/philic surfaces by Schwierz~\textit{et al.}~\cite{schwierz}.

\subsubsection{Desorption at different temperature}

Eventually, we examined the role of temperature $T$ in the process of chain
detachment and its impact on the force-displacement diagram. $T$ (measured in
units of the monomer - monomer interaction strength $\epsilon / k_B$) was
increased, while the adsorption strength $\epsilon_s$ was also correspondingly
changed so as to keep the ratio $\epsilon_s/k_BT$ constant and equal to
$\epsilon_s/k_BT = 20$ as in most of the presently studied cases. The presence
or absence of explicit solvent revealed thereby almost no difference again.

Expectedly, the mean level of the force (the average plateau height) grows,
reflecting the stronger adhesion $\epsilon_s$, while, surprisingly, the
amplitude of the spikes remains largely unchanged.
\begin{figure}[htb]
\begin{center}
\includegraphics[scale=0.25]{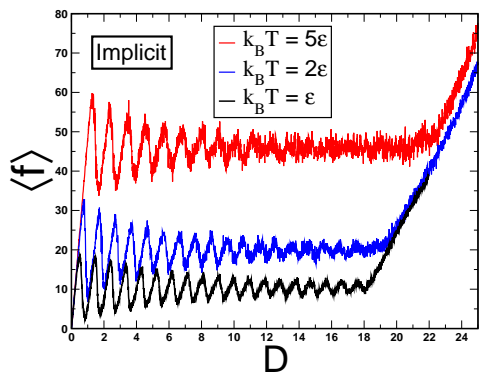}

\caption{ (a) Comparison of pulling results performed at different temperatures
$T$ (see legend) and constant ratio $\epsilon_s/k_BT = 20$ and $N=20$.
} \label{Temperature}
\end{center}
\end{figure}
A simple explanation for this observation in the force detachment experiment
can be suggested as follows.

We consider the chemical potential of adsorbed, $\mu_{\rm ads}$, and detached
segments, $\mu_{\rm det}$, which should become equal on the detachment line,
i.e., $\mu_{\rm ads} = \mu_{\rm det}$. In the limit of strong adsorption (or, at
low temperature), when the macromolecule is tightly bound to the surface and
loops (non-adsorbed chain portions) may be neglected, the free energy gain (per
chain segment) upon adsorption reads:
\begin{eqnarray}
  \mu_{\rm ads} = \underbrace{- \epsilon_s }_{\rm energy \:\: gain} -
  k_BT \underbrace{\ln\left(\mu_2/\mu_3\right)}_{\rm entropy \:\: loss}
\label{Mu_Ads}
\end{eqnarray}
In Eq.~(\ref{Mu_Ads}) $\epsilon_s$ again stands for the
adsorption energy of a single segment while $\mu_2$ and $\mu_3$ are the so
called {\it connective constants} in two- and three dimensional space
respectively. The latter correspond roughly to the possible orientations of a
chain segment in space, i.e., the logarithms thereof yielding effectively the
entropy contributions in two- and three-dimensions. It has been shown that for
cubic lattices, for instance, $\mu_2 = 2.6$ and $\mu_3 = 4.68$
\cite{Vanderzande}.

On the other  hand, in the limit of  strong adsorption, the detached chain
portion is strongly stretched, attaining a ``string'' configuration, so that the
elastic free energy per segment reads $\mu_{\rm det} = - a f$, where $a$ is the
Kuhn length and $f$ is the force acting on the chain end. Moreover, in the
``string'' state a segment has only one orientation , i.e. $\mu_3 = 1$. On the
$\langle f \rangle-D$ plateau, $f = f_{\rm p}$, and due to the condition
$\mu_{\rm ads} = \mu_{\rm det}$ one has the following ``plateau''-relationship
\begin{eqnarray}
 a f_{\rm p} = \epsilon_s + k_BT \ln \mu_2
\label{Plateau}
\end{eqnarray}
This result shows that for a strong adsorption the plateau height (i.e., the
pulling force) is proportional to the adsorption energy. The result given by Eq.
(\ref{Plateau}) has been obtained first within  a more general consideration in
our paper \cite{Paturej_MM} (see Eq. (30) in \cite{Paturej_MM}). This is now
supported by Figure~\ref{Temperature}  where the temperature and adsorption
energy are changed proportionally to one another.

The $\langle f \rangle -  D$ diagram demonstrates in all cases the
characteristic saw-tooth behavior with the amplitude progressively decaying in
the course of chain detachment. This behavior has been analyzed first by Jagota
{\it et al.} \cite{Jagota_1}. In terms of the number of detached chain segments,
the $n$-th spike correspond to the reversible transition $n \leftrightarrow n+1$
during which the detachment of a segment leads to release of polymer stretching
energy back to the energy of adsorption $\epsilon_s$. This condition leads to
the spikes amplitude law \cite{Jagota_1}
\begin{eqnarray}
f_{\rm amp} \sim \exp[(\epsilon_s/k_BT  - \ln 4\pi)/n].
\end{eqnarray}
This relationship is clearly in line with the decay behavior upon growing $n$.
On the other hand, provided that the temperature and adsorption energy are
increased proportionally to each other, the spikes amplitude does not change.
That is exactly what we observe in Figure \ref{Temperature}.

Figure \ref{Temperature} also shows that the overall elastic modulus of the
tethered chain does not depend on temperature. Most probably, this is due to the
fact that for the strong stretching the entropic contribution to elastic modulus
(modulus grows with temperature) is compensated by the bond anharmonicity effect
when the elastic modulus decreases with temperature.

\section{All-atom simulations} \label{sec_AllAtom}

\subsection{Model}
It appears instructive to compare the obtained simulation results for a
coarse-grained model to those for  an  atomistic model of a
concrete macromolecule. The latter were performed with the Gromacs MD
package~\cite{gromacs} using the Gromos96 force field~\cite{gromos96} and the
SPC/E (Single Point Charge/Extended) water model~\cite{spce} at  constant surface area $A$
and at constant vertical pressure $P_z$ of $1$~bar with temperature $T = 300$~K.
For the temperature and pressure control, the method of Berendsen~\cite{berendsen} was used. Periodic
boundary conditions for the Coulomb interactions were implemented by the
particle-mesh Ewald method~\cite{ewald}. Simulation runs were performed with an
integration time step equal to $2$~fs.
\begin{center}
\begin{figure}[ht]
\includegraphics[scale=0.35]{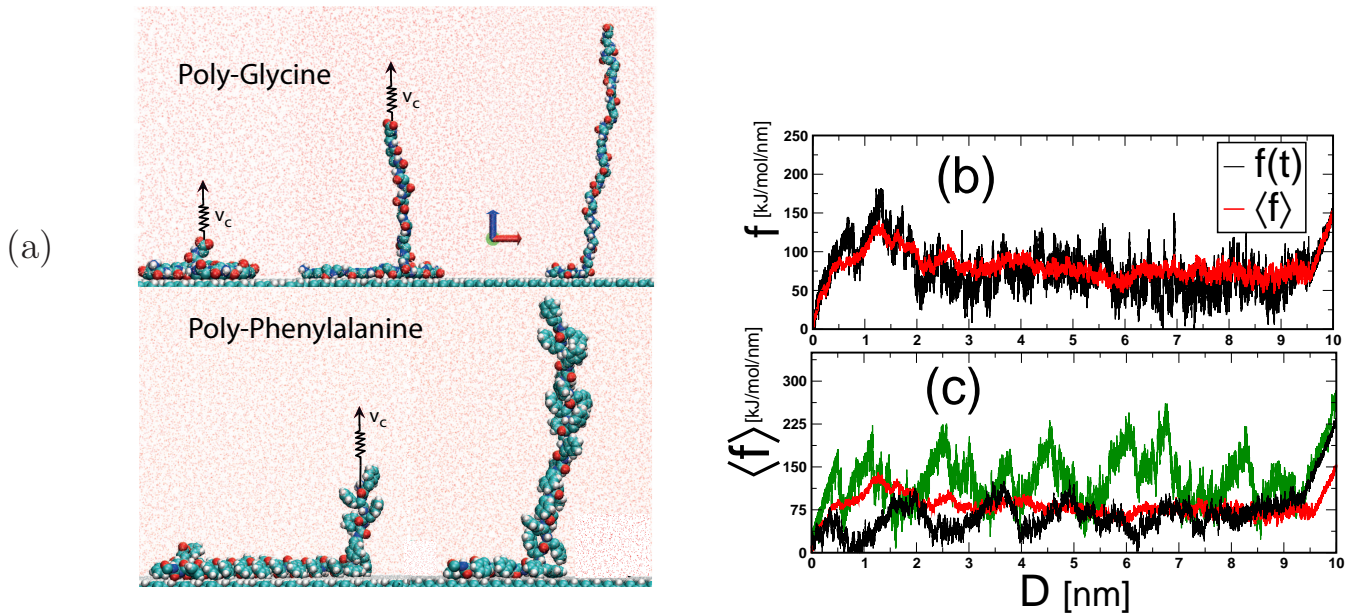}
\hspace{0.5cm}
\includegraphics[scale=0.32]{fvsD_glycin_N31.eps}
\begin{picture}(0,0)
 \put(-510,135){\Large (a)}
\end{picture}
\caption{{\it Biomolecule (polypeptide) desorption}. (a)
Vertical pulling of 31-glycine and 31-phenylalanine chains adsorbed on hydrophobic diamond surface.
Small red dots represent water molecules. (b) Force $\langle f \rangle$ vs $D$
diagram for $31$-glycine desorption. Red line represents an average over $25$
simulated desorption events whereas the back line show denotes a single run
simulation data.
(c) Averaged force $\langle f \rangle$ vs $D$ diagrams for: $31$-glycine (red line), $6$-(Gly-Gly-Gly-Gly-Phe) (black line) and $31$-phenylalanine (green line).
Here $k_c=200$, $\epsilon_s/k_BT \approx 7$ and the
pulling velocity $v_c=1$~m/s. }\label{glycine}
\end{figure}
\end{center}

The simulation box contains a hydrophobic diamond slab with a water-surface contact angle  $\theta_c \approx 90^o$~\cite{contactangle}, a single $N=31$  amino acid (AA) chain, and ca.~$16 000$  SPC/E water molecules. The entire system, including diamond surface, is
composed of $70 000$ atoms. Dimensions of the simulation box are around
$7$~nm$\times7$~nm$\times 12$~nm, and the thickness of a diamond slab is
$1.8$~nm. The ratio between adsorption strength of the surface and thermal energy is around $7-10$.
This binding energy is of the order of the binding energy $8.3 \pm 0.7$ of
polythymine $3'$poly(dC$_{50}$ on graphite substrate \cite{Jagota_2}. For the hydrophobic surface, the $\langle 100 \rangle$-plane of an
elastic  diamond substrate is saturated completely by uncharged hydrogen atoms.
The peptide is allowed to adsorb on the surface prior to solvation by water.
After equilibration of the chain on the surface, the pulling is performed, cf.
Figure~\ref{glycine}a, whereby initially the molecule is completely
adsorbed on surface.

Similar to CG simulations, the molecules were pulled
via a harmonic linker attached to one of the ends of the molecule as shown in Figure~\ref{glycine}. The  harmonic spring is  moved vertically at a prescribed velocity $v_c$
until the entire chain is desorbed from the surface. The spring exerts no
lateral force  on the chain, and the chain can move freely on the surface. The
spring constant is chosen as $k_c = 300$~pN/nm. The force needed to pull the
peptide vertically is calculated via $F= k_c(v_c t - z_t )$, where $z_t$ is the
$z$-component of the position of the terminal amino acid. The velocity is taken
as $v_c = 1$~m/s, which has been demonstrated earlier to provide a
quasi-equilibrium pulling on the corresponding surface~\cite{PNAS}.

Indeed, a pulling velocity $v_c \lesssim 1$ can also be justified by  scaling arguments  borrowed from the polymer physics: When HI is included, the relaxation time of chain with  $N$ Kuhn monomers is $\tau_Z \approx N^{3\nu} \tau_0$~\cite{Khokhlov}, where  the scaling exponent $\nu\approx3/5$ for good solvent, $\tau_0 \approx a^2 \gamma_0/k_{B} T$ is the  relaxation time of a Kuhn segment with a monomeric friction coefficient in the solvent $\gamma_0$.  If $\tau_Z$  is smaller than the time scale imposed by the pulling $\tau_c \approx a/v _c$, i.e.,  $v_c \lesssim v_c^* \equiv k_{B} T/a N^{3\nu} \gamma_0$, the undesorbed section of the chain will be in quasi-equilibrium for which pulling forces should not depend on the conformation of the remaining chain section on the surface.
If we take the Kuhn segment  size (twice the persistence length)  of an AA chain  as $a \approx 1$~nm (i.e.,  1-3 AA monomers),  $N \approx 10$ and  a monomeric friction coefficient of $\gamma_0 \approx 10^{-12}$~kg/s for the diamond surface~\cite{Erbas} and $k_{B} T \approx 4 \times 10^{-21}$kgm/s$^2$,  we obtain a threshold velocity $v_c^* \approx 1$~m/s.  Note that $v_c^*$ will be much lower if the monomeric friction coefficient is $\gamma_0 \gg10^{-12}$~kg/s, e.g. for a OH saturated surface~\cite{Erbas} or if the chains are longer.

Above argument for the chain relaxation on the atomistic surface also applieds to our CG simulations since in most of our CG simulations (except for the rough surface), the pulled chains interact with the surface only via a $z$-dependent potential. This means that chains can laterally   diffuse on the surface but with a bulk diffusion coefficient (see Section II.A).

\subsection{Comparison with Coarse-Grained simulations}

The all-atom simulations of polypeptide desorption from atomistically
rough substrate  were performed for a  $N=31$ polyglycine (31-glycine),  $N=31$ polyphenylalanine (31-phenylalanine). We also constructed a  $N=31$ hetero-peptide composed of six  (Gly-Gly-Gly-Gly-Phe) groups, where Gly and Phe stand for glycine and phenylalanine monomers, respectively.

The  results for 31-glycine chain desorption are presented in Figure~\ref{glycine}b along with several snapshots from different stages of the desorption event.
The snapshots taken from 31-glycine pulling trajectories compare well to those obtained from coarse-grained simulations shown in Fig.~\ref{snapshots}.
Comparing the first few spikes in Figure~\ref{glycine}b
and in Figure~\ref{Torsion}, the general pattern resembles much more the
saw-tooth profile, typical for semi-rigid, rather than for completely flexible
polymers, in line with the nature of this polypeptide.
Although the data for the $\langle f \rangle$-$D$ diagram, shown in
Figure~\ref{glycine}b, are averaged over $25$ simulation runs only and appear
somewhat noisy, they reveal a characteristic $\langle f \rangle - D$ behavior,
which qualitatively complies with the results from our coarse-grained
simulation. Due to the relatively
short length of this glycine macromolecule, the final 'dip' before contact with
the adhesive substrate is lost, is rather short yet clearly visible as in the
CG $\langle f \rangle$-$D$ diagrams of tethered chains.

To compare  31-glycine force trace with a stiffer chain, desorption simulations of 31-phenylalanine  (red data in Figure~\ref{glycine}c) were performed. As seen in  Figure~\ref{glycine}c, the saw-tooth peaks are more visible, and the peak-forces are much higher than those observed for 31-glycine cases.
This is actually due to the large benzyl side chain of phenylalanine monomers: The hydrophobic nature of the side chain  increases the affinity of  phenylalanine monomers to the hydrophobic surface, hence, results in higher force peaks.
 Interestingly, visual inspection of our simulation trajectories revealed that  the benzyl side chains  force the overall 31-phenylalanine molecule to take a rod-like structure on the surface (see the snapshot in Figure~\ref{glycine}a ). However, the pulling snapshots shown  in Fig.~\ref{glycine}a show that the conformation of 31-phenylalanine during the pulling resembles more  that of the 31-glycine rather the illustration shown in Fig.~\ref{snapshots} for the CG model with $\kappa \gg 25$. We attribute this to the similar atomistic AA backbone structure of both chains.
\\
\\
The $\langle f \rangle$-$D$ diagram of our  hetero-peptide  chain composed of (Gly-Gly-Gly-Gly-Phe) groups is shown in Figure~\ref{glycine}c (black data). One can  distinguish individual desorption peaks for 6 phenylalanine monomers separated by  peel-off's of glycine monomers which is also observed in CG simulations of alternating polymers (see Figure 5b).
The phenylalanine-induced force peaks  observed in the force trace of hetero-peptide chain are lower than those observed for  the 31-phenylalanine chain itself.   This 2-fold  difference in the peak forces can be due to complex interplay of chain stiffness and the relative surface affinity of monomers with respect to neighboring monomers: Possibly, glycine monomers might decrease the adsorption energy of adjacent phenylalanine monomers since they can diffuse faster due to  their relatively small sizes  (the side chain of a glycine is one hydrogen).   This observation in Fig.~\ref{glycine}c hints that the adsorption energy per AA residue can have a dependence on sequence and deserves further investigation in future.
Also note that the maxima of saw-tooth in Fig.~\ref{Copolymer}b  for CG model show a tendency to decrease as the adhesion asymmetry
of alternating monomers grows.
\\
\\
Overall, by comparing CG and atomistic simulations, one may conclude that the coarse-grained modelling of force-induced desorption of a polymer chain from adhesive substrate agrees well with the results from
all-atom simulations.

\section{Summary} \label{sec_Summary}

In the present investigation we studied the process of polymer chain detachment
by an external force, applied to the end-segment of a semiflexible chain which
is strongly adsorbed to adhesive substrate. Most of the results
have been derived by means of Molecular Dynamics simulations of a coarse-grained
bead-spring model of a polymer chain, and focused on the analysis of the
recorded (fluctuating) mean force $\langle f \rangle$ at height $D$ of the last
segment of the chain above the adsorbing plane when the segment is pulled with
given velocity $v_c$. As a principal objective of this
investigation, the
influence of different parameters that characterize the polymer chain, its
adhesion to the substrate, and the substrate - solvent affinity on the ensuing
$\langle f \rangle$-$D$ diagram have been examined.

We have found that an increasing bending rigidity of the polymer induces a sharp
drop of the pulling force {\it before} the last segments of the chain are peeled
off, i.e., the final portion of the chain is detached as a single piece of rod.
Nonetheless, our observations and estimates suggest that the sequential
desorption of polymer repeatable units from the substrate retains its
characteristic ``unzipping'' mechanism, reflected by the observed ``saw-tooth``
$\langle f \rangle - D$ profile, up to very high degree of rigidity. This
mechanism works not only for fully flexible chains but also for rather stiff
ones due to the gradual bending of the macromolecule which is energetically much
more favorable.

We also find that with increased bending stiffness $\kappa$, the modulation of
the characteristic oscillatory profile steadily declines, similar to the effect
of weaker attraction $\epsilon_s$ of the chain to the adsorbing surface where
the spikes vanish already at $\epsilon_s \approx 5 k_BT$. In contrast, the
torsional stiffness of the polymer has little or no effect of the $\langle f
\rangle$-$D$ diagram.

Regarding the possible use of the $\langle f \rangle$-$D$ diagram for sequencing
and its legibility, the performed detachment of an $A-B$-copolymer indicates
that the ratio $\epsilon_s^B / \epsilon_s^A$ of binding energies of the $A-$
and $B$-segments strongly influences the resulting oscillatory profile so that
when $\epsilon_s^B / \epsilon_s^A < 10\%$ the spikes that refer to $B$-atoms
practically disappear.

Our studies indicate that the role of hydrodynamic interactions (HI) in the
process of forced-induced detachment of a macromolecule from adsorbing surface
is negligible. The resulting $\langle f \rangle$-$D$ diagrams, emerging from MD
simulations with and without explicit solvent, hardly warrant the incomparably
larger computational efforts in the former case. This insensitivity of the
problem regarding HI is related most probably to the resulting stretched
conformation of the pulled macromolecule, and to the effect of screening of HI
in the vicinity of the adsorbing surface.

Eventually, by comparing our data derived from a coarse-grained bead-spring
model of a macromolecule to data from a realistic all-atom simulation of various
bio-polymer (e.g., glycine, phenylalanine), peeled off a hydrophobic diamond substrate, we have demonstrated that
the observed $\langle f \rangle$-$D$ diagrams agree qualitatively well with each
other, underlying thus the relevance of coarse-grained computer modelling.

\section*{Acknowledgments}
This research was supported
by the Polish Ministry of Science and Higher Education – grant {\it Iuventus Plus} Project No.: IP2012 005072.


\begin{thebibliography}{10}
\bibliographystyle{plain}

\bibitem{Merkel} R. Merkel, \emph{Phys. Rep.} \textbf{2001}, 346 , 343 .
\bibitem{Ritort} F. Ritort, \emph{J. Phys.: Condens. Matter} \textbf{2006}, 18, R531.
\bibitem{Franco} I. Franco, M.A. Ratner, G.C. Scatz, Single-Molecule Pulling:
Phenomenology and interpretation, in Nano and Cell Mechanics: Fundamentals and
Fronties, edited by H.D. Espinosa and G. Bao (Wiley, Microsystem and
Nanotechnology Series, 2013, ch. 14, pp. 359-388).
\bibitem{gromacs} Lindahl,~E.; Hess,~B.; van~der Spoel,~D. \emph{J. Mol.
Model.} \textbf{2001}, \emph{7}, 306317
\bibitem{Bell} G.I. Bell, \emph{Science} \textbf{1978}, 200, 618.
\bibitem{Evans_1} E. Evans, K. Ritchie, \emph{Biophys. J.}  \textbf{1997}, 72,
1541.
\bibitem{Evans_2}E. Evans,\emph{Annu. Rev. Biophys. Biomol. Struct.},
\textbf{2001}, 30, 105.
\bibitem{Baro} A.M. Bar\'o, R.G. Reifenberger (editors), Atomic Force
Microscopy in Liquid: Biological Applications, Wiley-VCH Verlag \& Co. KGaA,
Weinheim, 2012.
\bibitem{Merkel_1} R. Merkel, P. Nassoy, K. Ritchi, E. Evans, \emph{Nature}
\textbf{1999}, \emph{397}, 50 .
\bibitem{Jagota_1} S. Manohar, A. Jagota, \emph{Phys. Rev. E} \textbf{2010},
\emph{81}, 021805.
\bibitem{Jagota_2} S. Manohar, A.R. Manz, K.E. Bancroft, Ch-Y. Hui, A. Jagota,
D.V. Vezenov, \emph{Nano. Lett.} \textbf{2008}, \emph{8}, 4365.
\bibitem{Jagota_3} S. Iliafar, D.V. Vezenov, A. Jagota, \emph{Langmuir}
\textbf{2013}, \emph{29}, 1435.
\bibitem{Jagota_4}  S. Iliafar, K. Wagner, S. Manohar, A. Jagota, D.V.
Vezenov,\emph{J. Phys. Chem. C } \textbf{2012}, \emph{116}, 13896.
\bibitem{Paturej} J. Paturej, J.L.A. Dubbeldam, V.G. Rostiashvili, A. Milchev,
T. Vilgis, \emph{Soft Matter} \textbf{2014}, \emph{10}, 2785.
\bibitem{kg} K. Kremer and G.S. Grest, {\emph J. Chem. Phys.}  {\textbf 1990},
{\emph 92}, 5057
\bibitem{schneider} T. Schneider and E. Stoll, {\emph Phys Rev B} {\textbf
1978}, {\emph 17}, 1302
\bibitem{dpd} R.~D. Groot and P.~B. Warren, {\emph J. Chem. Phys.} {\textbf 1997}, {\emph 107}, 4423
\bibitem{frenkel} I. Pagonabarraga, M.~H.~J. Hagen and D. Frenkel, {\emph EPL} {\textbf 1998}, {\emph 42}, 377
\bibitem{soddemann} T. Sodemann, B. D\"unweg and K. Kremer, {\emph Phys. Rev. E} {\textbf 2003}, {\emph 68}, 046702
\bibitem{lammps} S.J. Plimpton, {\emph J. Comp. Phys.} {\textbf 1995} {\emph 117}, 1.
\bibitem{Khokhlov} A. Yu. Grosberg, A.R. Khokhlov, \emph{Statistical Physics of
Macromolecules}, AIP Press, New York, 1994.
\bibitem{gromos96} ~W.~R.~P. Scott,~P.~H.  H{\"u}nenberger,~I.~G. Tironi,~A.~E. Mark,~S.~R. Billeter,~J. Fennen,~A.~E. Torda,~T. Huber, ~P. Kr{\"u}ger,~W.~F  van Gunsteren, \emph{J. Phys. Chem. A} \textbf{1999},
\emph{103}, 3596.
\bibitem{spce} H. J. C. Berendsen, J. R. Grigera, and T. P. Straatsma, \emph{J. Phys. Chem}  \textbf{1987},  \emph{91}, 6269
\bibitem{berendsen} ,~K.~A. Feenstra,~B Hess,~H.~J.~S. Berendsen \emph{J. Comput. Chem.} \textbf{1999},   \emph{20}, 786.
\bibitem{ewald} T. Darden,~D. York, ~L. Pedersen \emph{J. Chem. Phys.} \textbf{1993},  \emph{98}, 10089
\bibitem{PNAS} D. Horinek,~A.  Serr,~M. Geisler,~T. Pirzer,~U. Slotta,~S.~Q.  Lud,~J.~A.  Garrido,~T.  Scheibel,~T. Hugel,~R.~R. Netz \emph{Proc. Natl. Acad. Sci.   U.S.A.} \textbf{2008}, \emph{105}, 2842
\bibitem{Winkler} A. Winkler, P. Virnau, K. Binder, R. G. Winkler, and G.
Gompper, \emph{Europhys. Lett.} \textbf{2012}, \emph{100}, 46003
\bibitem{Vanderzande} C. Vanderzande, {\it Lattice Model of Polymer}, Cambridge
University press, Cambridge, 2004.
\bibitem{Paturej_MM} J. Paturej, A. Milchev, V.G. Rostiashvili, T.A. Vilgis,
Macromolecules {\bf 45}, 4371 (2012).
\bibitem{contactangle} F.	Sedlmeier   \emph{1et al}. \emph{Biointerphases} \textbf{2008}, \emph{3},  23–39
\bibitem{Erbas} A. Erbas, D. Horinek,  R.R. Netz, \emph{J.  Am. Chem. Soc.} \textbf{2012}, \emph{134}, 623
\bibitem{schwierz}N.~Schwierz \emph{et al}. \emph{J.  Am. Chem. Soc.} \textbf{2012}, 134, 19628–19638


\end{thebibliography}
\end{document}